\documentclass[aps,pre,reprint,showpacs,amsmath,amssymb,floatfix,tightenlines]{revtex4-1} 

\usepackage[pdftex]{graphicx}       
\usepackage{txfonts}        
\usepackage{amsmath}
\usepackage{amssymb}
\usepackage{mathrsfs}	

\newcommand{\kcinf}{\kappa_\text{ads}}

\newcommand{\Tcinf}{T_\text{ads}}

\newcommand{\wTwoD}{\omega_\theta^\text{(2D)}}
\newcommand{\wThreeD}{\omega_\theta^\text{(3D)}}

\newcommand{\eref}[1]{Eq.~\eqref{#1}}
\newcommand{\fref}[1]{Fig.~\ref{#1}}
\newcommand{\sref}[1]{Section \ref{#1}}

\usepackage{hyperref} 
\hypersetup{colorlinks,citecolor=blue,filecolor=blue,linkcolor=blue,urlcolor=blue}

\begin{document}

\title{Phase transitions in solvent dependent polymer adsorption in three dimensions}

\author{C. J. Bradly} \email{chris.bradly@unimelb.edu.au}
\author{A. L. Owczarek}\email{owczarek@unimelb.edu.au}
\affiliation{School of Mathematics and Statistics, University of Melbourne, Victoria 3010, Australia}
\author{T. Prellberg} \email{t.prellberg@qmul.ac.uk}
\affiliation{School of Mathematical Sciences, Queen Mary University of London, Mile End Road, London, E1 4NS, United Kingdom}

\date{\today}
\begin{abstract}
We consider the phase diagram of self-avoiding walks (SAW) on the simple cubic lattice subject to surface and bulk interactions, modeling an adsorbing surface and variable solvent quality for a polymer in dilute solution, respectively.
We simulate SAWs at specific interaction strengths to focus on locating certain transitions and their critical behavior.
By collating these new results with previous results we sketch the complete phase diagram and show how the adsorption transition is affected by changing the bulk interaction strength. This expands on recent work considering how adsorption is affected by solvent quality. We demonstrate that changes in the adsorption crossover exponent coincide with phase boundaries. 
\end{abstract}

\maketitle

\section{Introduction}
\label{sec:Intro}

Adsorption of polymers in dilute solution and the associated critical behavior is a long standing topic in statistical physics \cite{Eisenriegler1982,DeBell1993,Vrbova1996,Vrbova1998,Vrbova1999,Grassberger2005,Owczarek2007,Luo2008,Klushin2013}.
The canonical model for such polymers is the self-avoiding walk (SAW) on a regular lattice, allowing for mean field-theoretic analysis \cite{DeGennes1975,Stephen1975,Duplantier1982} as well as extensive numerical simulation \cite{Rapaport1985,Madras1988}.
At high temperatures, a polymer in a good solvent forms an extended coil configuration, seeking to maximize entropy.
Below a certain temperature $\Tcinf$ it is energetically favorable for the polymer to be adsorbed to an attractive surface where the fraction of the polymer lying on the surface approaches unity.
Solvent quality is another influence on the conformational properties of polymers in dilute solution and is modeled by a monomer-monomer interaction.
Polymers modeled by SAWs are therefore an important model for considering the interplay between surface and bulk interactions.

The relevant order parameter for adsorption of polymers is the fraction of the polymer lying on the surface
\begin{equation}
	u_n  = \frac{\langle a \rangle}{n} \sim n^{\phi-1},
	\label{eq:AverageContacts}
\end{equation}
where $a$ is the number of monomers adsorbed to the surface, $n$ is the length of the polymer chain and the scaling is determined by the exponent $\phi$.
Clearly, $\phi=1$ in fully adsorbed phases but takes on other values in other phases and at the transitions between phases.
In a good solvent, $\phi$ becomes a crossover exponent at the critical temperature $\Tcinf$ controlling critical behavior and it has been proposed \cite{Hegger1994,Metzger2003} that $\phi=1/2$ for SAWs in any dimension, making $\phi$ superuniversal at the adsorption transition.
However, recent consensus due to numerical simulation is that $\phi$ is not super universal
\cite{Grassberger2005,Klushin2013,Luo2008,Bradly2018}.
Numerical simulation is thus a useful tool in this field and can be applied to other questions.

For the effect of the bulk interaction on polymer adsorption, recent work by Plascak {\em et al} \cite{Plascak2017,Martins2018} has suggested that altering the strength of the bulk interactions with respect to the surface interactions has a significant effect on $\phi$ and the critical temperature $\Tcinf$.
For the case where the bulk interaction is made increasingly repulsive the critical temperature decreases slightly from the noninteracting case.
While Ref.~\cite{Plascak2017} claims that $\phi$ also decreases slightly in the same limit, we have used variations on the self-avoiding walk to mimic strongly repulsive bulk interactions, finding no good evidence that $\phi$ changes in this limit \cite{Bradly2018b}.

The complete phase diagram of SAWs with both surface and bulk interactions has been extensively studied with numerical simulations \cite{Bachmann2005,Bachmann2006,Martins2016} and exact enumeration \cite{Mishra2005,Singh2001} but some details remain in doubt.
There are many phases and it can be hard to isolate particular transitions due to finite-size effects \cite{Bradly2018} and difficulties with determining the correct signature of the transition \cite{Qi2018}.
In this article we look at the entire phase diagram and present results of new simulations so as to focus on attractive bulk interactions.
The large changes to $\phi$ and $\Tcinf$ as the bulk interaction strength is increased coincide with the appearance of other phases where bulk collapse is just as important as surface adsorption and the critical temperatures are not $\Tcinf$ but are indicative of a transition to these other phases. This implies a simpler picture of the variation of  $\phi$ with bulk interaction strength being constant on phase boundaries between similar phases and discontinuously jumping when the phase transition changes type: that is, the normal universal behaviour of exponents.
Overall, we are able to locate these other phase boundaries and thus sketch out the entire phase diagram.

\section{Model and phase diagram}
\label{sec:Model}

Single polymers are modeled as self-avoiding walks (SAWs) on the positive half-space of the simple cubic lattice.
The canonical partition function for walks of length $n$ with $m$ bulk interactions and $a$ surface interactions is
\begin{equation}
	Z_n(T) = \sum_{a,m} c_{n}(a,m) \exp\left( \frac{a \epsilon_\text{surf} + m \epsilon_\text{bulk}}{k_\text{B}T} \right),
	\label{eq:CombinedPartition}
\end{equation}
where $c_{n}(a,m)$ is the density of states, $-\epsilon_\text{surf}$ is the interaction energy of a point in the SAW in contact with the surface and $-\epsilon_\text{bulk}$ is the interaction energy of a pair of nonconsecutive steps of the SAW on neighboring lattice points.
There is some freedom in how to parametrize the thermal or Boltzmann factor in \eref{eq:CombinedPartition} depending on perspective and preference.
In Ref.~\cite{Plascak2017} the authors use temperature $T$ and energy ratio $s=\epsilon_\text{bulk}/\epsilon_\text{surf}$.
In particular, $\epsilon_\text{surf}$ defines the energy scale so that large $s$ means bulk interactions are energetically favorable, small $s$ means surface interactions are energetically favorable and negative $s$ means bulk interactions are repulsive.

As an alternative, in our previous work we assign separate Boltzmann factors to the bulk and surface interactions so that the total thermal factor is written $\kappa^a\omega^m$, where $\kappa = \exp(\epsilon_\text{surf}/k_\text{B}T)$ and $\omega = \exp(\epsilon_\text{bulk}/k_\text{B}T)$.
In this parametrization the energetically favored interaction is determined by the larger of $\kappa$ and $\omega$ and repulsive interactions are represented by $\kappa<1$ or $\omega<1$.

\begin{figure}[t!]
	\centering
	\includegraphics[width=\columnwidth]{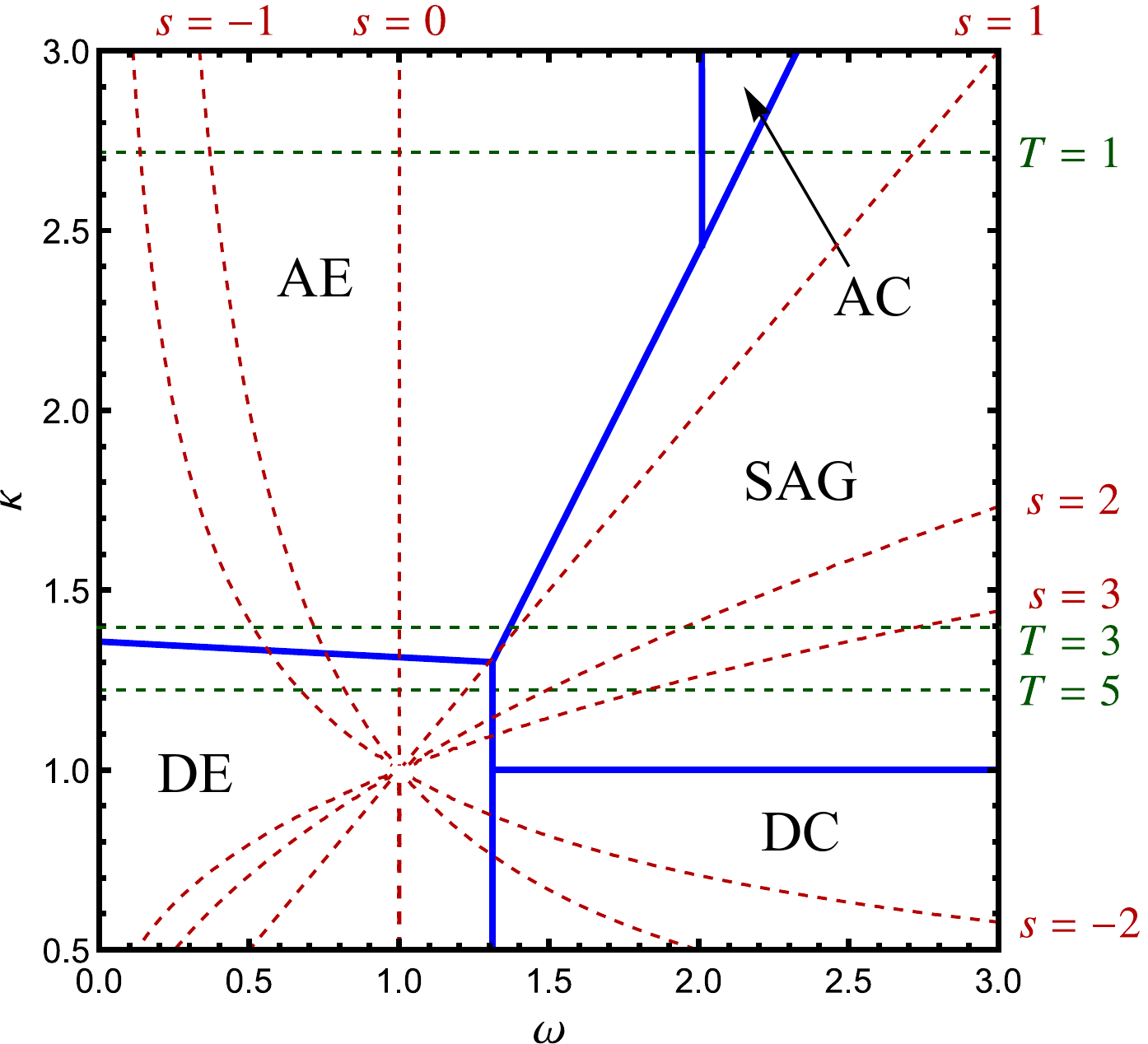}
	\caption{Schematic phase diagram for adsorbing and interacting SAWs in three-dimensions based on known critical points and exponents. 
	The phase diagram is in the $\omega$-$\kappa$ parametrization and is overlaid with contours of fixed $s$ and $T$ (dashed lines)}
	\label{fig:SchematicPhase}
\end{figure}

In the $\kappa$-$\omega$ picture, we show in \fref{fig:SchematicPhase} the schematic phase diagram for adsorbing and interacting SAWs in three-dimensions based on known results.
The dashed lines show how this picture maps on to the $s$-$T$ parameterisation and the blue lines demarcate the phases by joining the known values of the critical points.
Starting from the pure-entropic desorbed-extended (DE) phase at high temperature, in the case of a non-interacting or repulsive surface ($\kappa \leq 1$) the polymer may undergo a $\theta$-point transition to the desorbed-collapsed (DC) phase, which for the simple cubic lattice occurs at $\wThreeD=1.31$ \cite{Grassberger1997}.
In both desorbed phases, the average surface fraction is trivially zero, represented by $\phi=0$.
Conversely, in the case of zero or repulsive bulk interactions ($\omega<\wThreeD$), a polymer in the DE phase may undergo the adsorption transition to the adsorbed-extended (AE) phase.
At the point of no bulk interactions ($\omega=1$), the critical temperature $\Tcinf$ is equivalent to $\kcinf=1.33$ \cite{Grassberger2005}, but there is a small shift in $\kcinf$ as $\omega$ is varied below $\wThreeD$.
For polymers fully adsorbed to the surface $\phi=1$, but at the adsorption transition it is expected that $\phi=1/2$, from mean-field predictions.
The three-dimensional case is slightly different with numerical evidence suggesting a slight deviation from the mean-field value; recent Monte Carlo studies converge near Grassberger's value $\phi\approx 0.48$ \cite{Grassberger2005,Klushin2013,Plascak2017,Bradly2018} but other values are possible \cite{Taylor2014}.
Further, there is some disagreement over the value of $\phi$ as the bulk interaction is changed from non-interacting ($\omega=1$) to strongly repulsive ($\omega=0$) \cite{Bradly2018b}.

As $\kappa\to\infty$ the model changes to a two-dimensional one, and the polymer may undergo collapse to a two-dimensional adsorbed-collapsed (AC) phase as $\omega$ increases.
The square lattice is the two-dimensional limit of the simple cubic lattice and the collapse occurs at $\wTwoD=1.94$ \cite{Meirovitch1989b}.
The collapse transition in two dimensions is weaker than in three dimensions and occurs at larger $\omega$, for those models where the same bulk interaction can be used.

Finally, there is another three-dimensional phase where bulk and surface interactions are strong and the polymer has the configuration of a surface-attached globule (SAG).
Comparing the volume of the globule to the fraction of its surface area that rests on the interacting surface suggests that $\phi=2/3$ in this phase \cite{Owczarek2007}.
In the limit of large $\omega$ and large $n$ the SAG-DC transition is expected to occur at $\kappa=1$, but there is still uncertainty as to where this boundary connects to the 3D collapse transition.
However, this is not the main focus of the current article.

The outstanding question in the study of adsorbing lattice polymers in three dimensions is what happens to the adsorption transition when the bulk interaction is attractive, $\omega > 1$.
As $\omega$ is increased the pure-adsorption AE-DE transition meets the pure-collapse DE-DC transition and the SAG-AE transition at a multicritical point, where $\phi$ is believed to return to its mean-field value of $1/2$.
The location and nature of the SAG-AE boundary is less well understood.
Precise determination of the critical temperature along this boundary is hindered by a number of factors.
Since the transition is both a surface and bulk transition, it is not obvious what is the best signature of the transition.
The methods we have explored for the adsorption transition \cite{Bradly2018} do not work as well here.
The complexity of the phase diagram limits the range of parameters that are sure to hit only the SAG-AE boundary without probing other transitions in the system.
This is further complicated by the appearance of a series of layering transitions in the weak solvent regime $\omega>\wThreeD$ \cite{Singh2001,Krawczyk2005,Krawczyk2005a}. 
These transitions are omitted in \fref{fig:SchematicPhase} since they only appear in finite-size systems but are a concern in numerical simulations.
Further, the location of the boundaries is dependent on $n$ in finite-size simulations so trying to calculate scaling of thermodynamic quantities over a range of $\kappa$ and $\omega$ is difficult.
Nevertheless, we can make some progress towards mapping out the missing parts of the phase diagram and confirming some signature properties of the phases, even if we do not have a high degree of numerical precision for some quantities.

\begin{figure}[t!]
	\centering
	\includegraphics[width=\columnwidth]{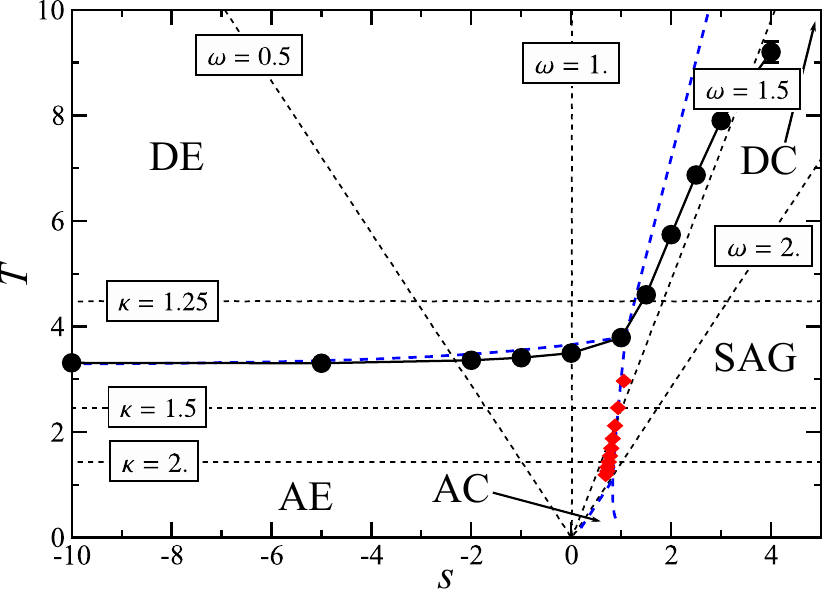}
	\caption{Phase diagram of Ref.~\cite{Plascak2017} (black circles) overlaid with schematic phase boundaries of Fig.~1 transformed to $s$-$T$ parametrization (blue dotted), contours of fixed $\omega$ (dotted rays) and $\kappa$ (dashed horizontal). 
	Red squares mark the critical temperatures from our fixed $\kappa$ simulations and mark the SAG-AE transition.}
	\label{fig:sTPhaseEverything}
\end{figure}

\section{Simulations and Results}
\label{sec:Simulation}
	
Walks are simulated using the flatPERM algorithm \cite{Prellberg2004}, an extension of the pruned and enriched Rosenbluth method (PERM) \cite{Grassberger1997}. 
We have used this method previously to study the adsorption transition without bulk interactions \cite{Bradly2018,Bradly2018b}.
The simulation works by growing a walk on a given lattice up to some maximum length $N_\text{max}$. 
At each step the number of bulk interactions $m$ and surface contacts $a$ are calculated and the cumulative Rosenbluth \& Rosenbluth weight \cite{Rosenbluth1955} is compared with the current estimate of the density of states $W_{n,m,a}$. 
If the current state has relatively low weight the walk is `pruned' back to an earlier state. 
On the other hand, if the current state has relatively high weight, then microcanonical quantities $m$ and $a$ are measured and $W_{n,m,a}$ is updated. 
The state is then `enriched' by branching the simulation into several possible further paths (which are explored when the current path is eventually pruned back). 
When all branches are pruned a new iteration is started from the origin.
FlatPERM enhances this method by altering the prune or enrich choice such that the sample histogram is flat in the microcanonical parameters $n$, $m$ and $a$. 
Further improvements are made to account for the correlation between branches that are grown from the same enrichment point, which provides an estimate of the number of effectively independent samples. 
We also run 10 completely independent simulations for each case to estimate the statistical error.
The main output of the simulation is the density of states $W_{n,m,a}$ which is an approximation to the athermal density of states $c_{n}(a,m)$ in \eref{eq:CombinedPartition}, for all \smash{$n\le N_\text{max}$}. 
In practice, thermodynamic quantities are determined by specifying $\kappa$ and $\omega$  and using the weighted sum
\begin{equation}
    \langle Q \rangle(\kappa,\omega) = \frac{\sum_{m,a} Q_{m,a}\omega^m\kappa^a W_{n,m,a}}{\sum_{m,a} \omega^m\kappa^a W_{n,m,a}}.
    \label{eq:FPQuantity}
\end{equation}

Producing flat histograms over both $a$ and $m$ simultaneously limits the maximum length $n$ that can be simulated.
Alternatively, we can fix one of the weights, $\kappa$ or $\omega$, within the simulation by including it as a constant factor in the total weight of the sample at all growth steps and only flatten over the other microcanonical parameter.
Fixing $\kappa$ or $\omega$ in this fashion is equivalent to simulating along horizontal or vertical lines in the phase diagram, respectively, and allows much longer walks to be simulated.

To map out the SAG-AE transition we first performed flatPERM simulations of SAWs on the cubic lattice up to length $n=1024$ at fixed values of $\kappa$ in the range $1.4\leq\kappa\leq 2.3$ for a total of ten independent simulations, each producing an average of $2.7\times 10^{11}$ samples.
This range of $\kappa$ avoids complications from where the multicritical points and other phases are expected to be.
The transition is a bulk transition as well as a surface transition, so we are not constrained to look for a specific signature of the transition like in the case of the adsorption transition. 
We estimate the location of the critical point by the peak of the variance of the order parameter
\begin{equation}
	\frac{\text{var}(m)}{n}=\frac{\langle m^2\rangle-\langle m\rangle^2}{n},
\label{eq:Variance}
\end{equation} 
as a function of $\omega$ and over a range of $n$.
The transition appears as a slightly broadened peak in the variance of the microcanonical parameters but this approach allows data to be collected at longer lengths.
The positions of the peaks for the range of $n$ are extrapolated to infinite lengths assuming a power law in accordance with standard finite-size scaling theory.
This process is repeated for each value of $\kappa$ to obtain a set of $(\omega,\kappa)$ pairs marking the SAG-AE boundary.
This method is deliberately simple, partly due to the complications addressed in \sref{sec:Model} and partly because our chief interest is to demarcate the SAG-AE boundary in contrast to other features of the phase diagram and not to obtain precise estimates of the transition temperature.

%

With these estimates of the location of the SAG-AE transition as $(\omega,\kappa)$ pairs, we return to the question of parametrization of the phase diagram.
In \fref{fig:sTPhaseEverything} we show the $s$-$T$ phase diagram from Ref.~\cite{Plascak2017} (black circles) augmented with our estimates of the SAG-AE transition transformed to the same parameters (red squares).
The blue dotted lines are the transformed schematic phase boundaries from \fref{fig:SchematicPhase}.
Also shown are contours of fixed $\omega$ (black dotted) and $\kappa$ (black dashed) to illustrate the transformation between parametrization and the regions covered by each study.

It is immediately clear that results for $s>1$ are actually showing the SAG-DE transition, which is really a collapse transition more like the $\theta$ transition, rather than a true surface transition in the presence of weak bulk interactions.
In fact the weakly attractive bulk interaction regime, found at $1<\omega<\omega_\theta\approx1.31$ in our parametrization, is contained entirely within $0<s\lesssim 1$, since the multicritical point is, coincidentally, very near $s=1$.
In contrast, The SAG-AE transition, which is a surface transition as well as a bulk transition, does occur near $s=1$ but at smaller $T$.

In addition to the critical temperatures, Ref.~\cite{Plascak2017} observed a large change to the value of $\phi$ in the $s>1$ regime, attributed to effects due to multicritical points. 
We can now see that the $s>1$ regime does not contain any adsorption transitions.
We also note that the conjectured boundary between the SAG and DC phases is indicated in \fref{fig:sTPhaseEverything} by large $s$ and large $T$ but the DC phase is not well-defined in this parametrization.
For finite systems there may be some observable effect due to the appearance of the DC phase at finite $T$ and a possible multicritical point at large $s$, but we do not expect this to be significant in the energy ranges considered here.
Therefore there are no further multicritical points to consider for $s>1$.
Instead, we expect the value of $\phi$ to be entirely dependent on the expected configurations of the SAG and DE phases and not display any critical behavior.
We can now see that this is the normal behavior of the system in the SAG and DE phases and the transition between them.


It is also apparent from \fref{fig:sTPhaseEverything} that the regime of weakly attractive bulk interactions, known to be between $\omega=1$ and $\omega=\omega_\theta$, is contained entirely within $0<s<1$, and as such has not yet been investigated.
This regime is of interest for matching the noninteracting and repulsive bulk interaction regimes, where we know that $\phi\approx0.48$, to the expected mean-field value $\phi=1/2$ at the multicritical point.
For other values of $s$ the phase diagram is better understood.
Where $s\leq 0$ (or $\omega\leq1$) the system is interpreted to have zero or repulsive bulk interactions, since the surface interaction energy $\epsilon_\text{surf}$ is used as a reference.
As the bulk interaction becomes more repulsive the critical temperature decreases slightly.
While Ref.~\cite{Plascak2017} found evidence that $\phi$ decreases in this regime, a study of neighbor-avoiding walks that model the infinitely repulsive limit imply that this could be within statistical and numerical error \cite{Bradly2018b}.
Otherwise, the adsorption transition is reasonably well understood in the case that the bulk interaction is below the collapse point.


\begin{figure}[t!]
	\centering
	\begin{tabular}{cc}
	\includegraphics[width=0.48\linewidth]{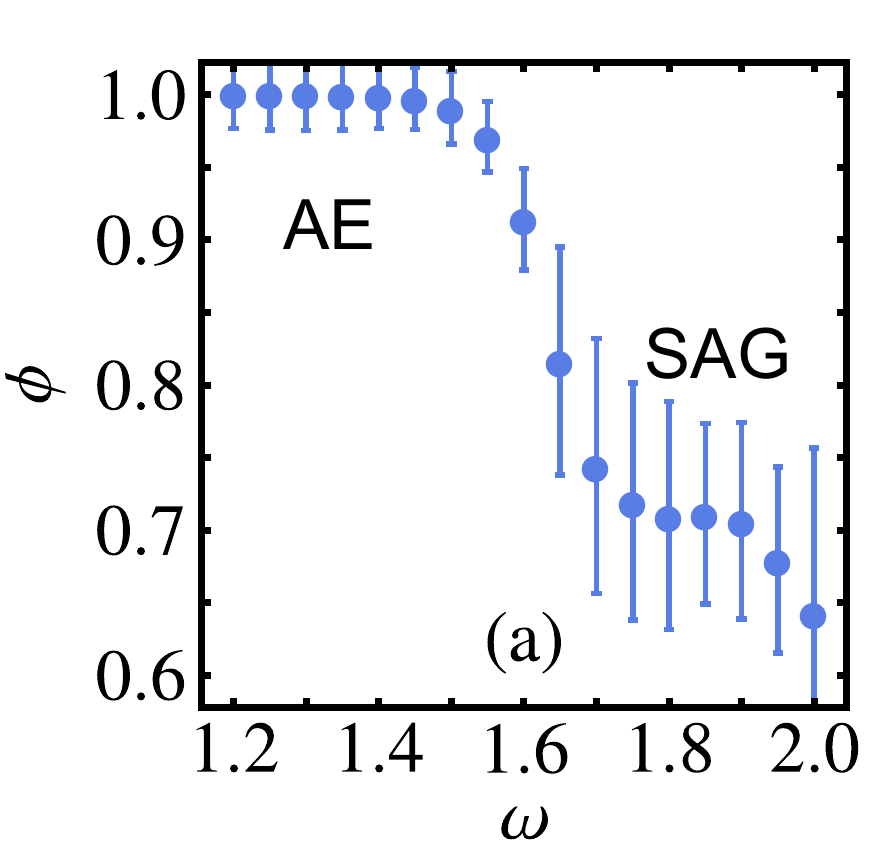}
	\label{fig:PhiHorizontal}
	&
	\includegraphics[width=0.48\linewidth]{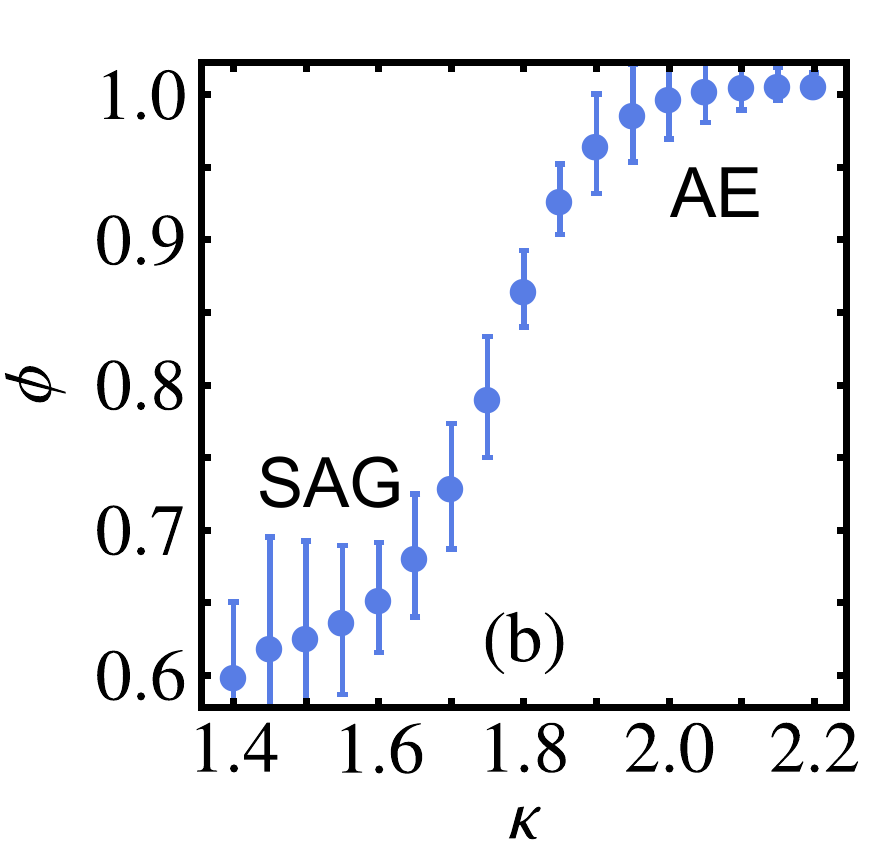}
	\label{fig:PhiVertical}
	\end{tabular}
	\caption{Exponent $\phi$ at (a) fixed $\kappa=1.8$ and (b) fixed $\omega=1.6$, across the SAG-AE boundary. In both cases results are consistent with $\phi=1$ for the adsorbed phase and $\phi=2/3$ for the surface-attached globule phase.	}
	\label{fig:SAG-AE_Phi}
\end{figure}

Our estimates of the critical point shown in \fref{fig:sTPhaseEverything} are a clear signature of a transition occurring, but are not accurate enough for determining other properties.
This is because in order to cover the whole SAG-AE boundary each point is from only a single simulation instance and only considers the variance of a single parameter.
However, having established the location of the phase boundaries we can focus on a specific point in the phase diagram near the SAG-AE boundary in order to investigate critical behavior.

For that purpose we ran two further simulations of SAWs on the simple cubic lattice up to length 1024 at fixed weights $\kappa=1.8$ and $\omega=1.6$, corresponding to a horizontal and vertical slice in the phase diagram, respectively.
Each of these simulations combined ten independent instances for $7.1\times 10^{12}$ and $9.6\times 10^{12}$ total samples, respectively.
This provides greater accuracy, particularly regarding the scaling of the order parameter $\langle a \rangle$ with length $n$.
The value of the exponent $\phi$ is determined from $\langle a \rangle$ by adding a correction-to-scaling term to \eref{eq:AverageContacts}, i.~e.~
\begin{equation}
    u_n=\frac{\langle a \rangle}{n} \sim n^{\phi-1} f^\text{(0)}(x) [1 + n^{-\Delta}f^\text{(1)}(x) +\ldots ],
    \label{eq:UnScaling}
\end{equation}
where the $f^{(i)}$ are  finite-size scaling functions of the scaling variable $x=(\Tcinf-T)\,n^{\phi}$ and so are assumed to be constant near the transition.
The exponent $\Delta$ determines the first correction-to-scaling term but its precise value has little effect provided $\Delta\lesssim 1$.
Figure \ref{fig:SAG-AE_Phi} shows the exponent $\phi$ across the SAG-AE boundary in two ways: (a) a horizontal slice at fixed $\kappa=1.8$ over a range of $\omega$ and (b) along a vertical slice at fixed $\omega=1.6$ over a range of $\kappa$. 
The intersection of these slices is near the SAG-AE boundary so with respect to this point, in the AE phase at smaller $\omega$ and larger $\kappa$ the exponent is $\phi=1$, consistent with the walk being fully adsorbed to the surface.
For larger $\omega$ and smaller $\kappa$, the SAG phase, our data is consistent with $\phi=2/3$.
Despite the increased focus on a single value of $\kappa$ and $\omega$ we are unable to determine precisely where the critical point is with enough accuracy to determine if $\phi=2/3$, $\phi=1$ or some intermediate value at the transition.
A dedicated study would be required to resolve this question.

\section{Conclusion}
\label{sec:Conclusion}

In this article we have resolved outstanding issues in the phase diagram of SAWs with bulk and surface interactions. 
The transition from the DE phase to the DC phase is independent of the surface interaction strength, and similarly, the bulk interaction strength only weakly affects the location of the transition to the AE phase and probably does not effect the critical behavior, namely the exponent $\phi$.
The collapse transition between the AE and AC phases, representing adsorption in two dimensions, occurs at higher values of the bulk and surface interactions, meaning there is an additional phase boundary joining the multicritical points. 
We have mapped this boundary by varying the bulk interaction at fixed values of the surface interaction.
While finite-size effects inhibit our ability to obtain highly accurate estimates of thermodynamic parameters in each phase, we are able to show how the exponent $\phi$ which controls the scaling of the order parameter varies across this boundary.
The behavior of $\phi$ is consistent with the presence of a SAG phase and not due to multicritical scaling as the system moves along the line of adsorption transition points to the multicritical point where collapse occurs, as suggested by Ref.~\cite{Plascak2017}.

All phases in this system have been identified but there are some remaining questions about the details that we have not yet addressed.
Having distinguished the SAG-AE transition as the bulk interaction is varied, future work can focus on the multicritical points or on deeper analysis of each transition.
The region near $\wThreeD$ for $1<\kappa<\kcinf$ requires careful attention to resolve where the DC-SAG transition joins the other phase boundaries, particularly comparing finite $n$ to infinitely long chains.
Understanding this transition will allow greater focus to be put on the SAG-AE transition in order to locate the phase boundary with enough accuracy to understand the critical behavior of this transition.

\begin{acknowledgments}
C.~B.~thanks Queen Mary University London for hosting while some of this work was carried out.
Financial support from the Australian Research Council via its Discovery Projects scheme (DP160103562) is gratefully acknowledged by the authors. 
\end{acknowledgments}

\bibliography{../../Papers/polymers_master}{}

\end{document}